# Instability of counter-rotating stellar disks

R G Hohlfeld[1,2,4] and R V E Lovelace[3]
[1]Center for Computational Science, Boston University, Boston, MA 02215 USA
[2]Wavelet Technologies, Inc., Attleroro, MA 02703 USA
[3]Astronomy Department, Cornell University, Ithaca NY USA

E-mail: hohlfeld@bu.edu

**Abstract.** We use an N-body simulation, constructed using GADGET-2, to investigate an accretion flow onto an astrophysical disk that is in the opposite sense to the disk's rotation. In order to separate dynamics intrinsic to the counter-rotating flow from the impact of the flow onto the disk, we consider an initial condition in which the counter-rotating flow is in an annular region immediately exterior the main portion of the astrophysical disk. Such counter-rotating flows are seen in systems such as NGC 4826 (known as the "Evil Eye Galaxy"). Interaction between the rotating and counter-rotating components is due to two-stream instability in the boundary region. A multi-armed spiral density wave is excited in the astrophysical disk and a density distribution with high azimuthal mode number is excited in the counter-rotating flow. Density fluctuations in the counter-rotating flow aggregate into larger clumps and some of the material in the counter-rotating flow is scattered to large radii. Accretion flow processes such as this are increasingly seen to be of importance in the evolution of multi-component galactic disks.

## 1. Introduction

Most disk galaxies, including our own Milky Way, have vast majority most of their stars and gas on orbits (hereafter rotating) with the same sense of rotation about the center of the galaxy. However, there is a significant population of disk galaxies in which there are substantial components that orbit in the opposite sense (or at high inclination) to the general orbital motion of their host galaxy. Detailed studies of the inner structures of galaxies are required to clearly identify counter-rotating flows. Several of the well-observed cases of this phenomenon include NGC 4826 (also known as M64, and colloquially as the "Evil Eye Galaxy" or the "Black Eye Galaxy" due to prominent dust lanes associated with the counter-rotating flow, see figure 1), NGC 7217, NGC 4550, and NGC 5150 [1, 2, 3]. These examples in neighboring galaxies, where detailed studies of the inner structure required to clearly identify counter-rotating structures, make plausible a statistical argument that galaxies with such prominent counter-rotating structures are not uncommon. This motivates research to understand how these systems originate and the important dynamical instabilities controlling their evolution.

---

[4] Author to whom any correspondence should be addressed.

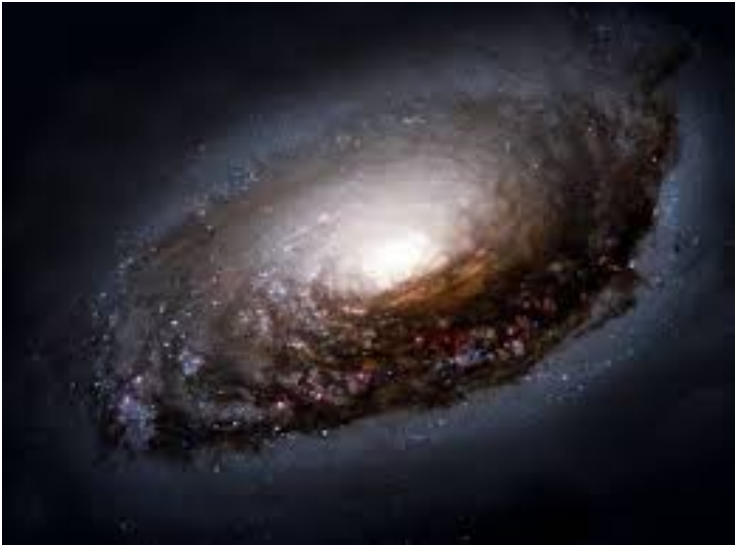

**Figure 1.** A view of NGC 4826. The boundary between counter-rotating and direct rotation is indicated by the dark dust lanes, approximately 3000 light years from the galaxy's center. Relative velocity of the two flows is 300 km/s.

Image Credit: NASA/ESA and the Hubble Heritage Team (AURA/STScI)

Much work has been done in accretion flows onto black holes, neutron stars, and other condensed objects in which the accreting material is essentially or entirely gaseous. It is reasonable to consider the accretion as resulting from the dynamics of ordinary viscosity in the gas. We are considering a more complex scenario in this proposal in which the entire galaxies under study have stellar as well as gaseous material, and so it is necessary to consider instability mechanisms deriving from both particle (i.e., stellar and cluster dynamics and fluid flow mechanisms. With a galactic accretion flow composed of counter-rotating stellar flows interpenetrating, or adjacent with a bounding shear layer, it is most attractive to model the interaction as a two-stream instability [4]. These theoretical studies of galactic accretion disk flows concentrated on relatively low azimuthal mode numbers. The simulation study reported here indicates that, at least for disks with relatively flat rotation curves, the azimuthal mode number for well-developed instabilities is comparatively high, i.e. $m \geq 6$. This point will be discussed in more detail below, but the major thrust of the proposed research will be to extend theory and simulations to a common range of parameters and investigate correspondence with observational results.

## 2. Tools and Methodology
We describe here the computational tools used in carrying out and analysing the simulation reported here.

### 2.1. *GADGET*
This study, and most of our other simulation studies, are conducted using GADGET, developed by Springel and coworkers [5, 6]. GADGET has been used extensively for simulations of galaxy collisions and interactions, and thus is well-suited for our studies of galaxy accretion and associated evolution.

GADGET is a Langrangian code, which enables the treatment of processes spanning a large range of spatial and time scales. Numerous particle types are treated in a natural way, including hydrodynamic particles treated by smoothed particle hydrodynamic (SPH) techniques. This is important for our purposes because we wish to address cases in which instabilities involving stellar particles and instabilities involving gas occur.

GADGET has been implemented in parallel versions using a Message Passing Interface (MPI) suitable for use on cluster computers. Such implementations allow ambitious simulations with large numbers of particles, that we have found useful for our work.

*2.2. Snapshot file analysis tools*

In GADGET input and output is typically handled by "snapshot" files, defined in the GADGET user guide [7]. Many workers have written programs for creating and processing snapshot files and we have similarly adapted software for processing snapshot files for our needs.

Our snapshot file analysis programs are written in *Mathematica 10*, which we have found to be powerful for its integrated visualization capabilities as well as smoothly tying in with our theoretical analysis tools.[5] Statistical functions in Mathematica are useful for the estimation of particle distribution functions, which is important for identifying the operation of instabilities driven by velocity space effects. Furthermore, we have implemented modal analysis tools suitable for determining the azimuthal mode spectrum of instabilities.

Our analysis tools make it a straightforward task to trace both the stellar and interstellar gas populations of particles from any sources, as well as different populations of particles, which is necessary for distinguishing particles originating from an accreting body relative to the particles in the accreting galaxy.

## 3. Initial Conditions for the Simulation

The simulation described here is designed to demonstrate the operation of two-stream instability in a accretion in a galactic disk made of stars. Accordingly, we use GADGET particles and do not make use of GADGET's capabilities for simulations using smoothed particle hydrodynamics. Future work will address gaseous systems and systems with mixtures of gas and stars. A total of 500,000 particles was used to obtain adequate resolution of features in the accretion flow.

These simulations used relatively flat (constant velocity) rotation curves because of our interest in Mestel disks and in large disk galaxies [8, 9]. We used a modification of the standard Mestel disk to avoid some of the singularities and related numerical problems associated with this equilibrium solution. Specifically, we pick a mass profile and rotation curve of the form

$$\sigma(r) = \frac{\sigma_0}{\sqrt{1+r^2/\lambda^2}} \qquad (1)$$

$$v_c^2(r) = 2\pi G \lambda \sigma_0 \left[1 - \frac{1}{\sqrt{1+r^2/\lambda^2}}\right] \qquad (2)$$

where σ is the mass surface density, $\sigma_0$ is the mass surface density at radius $r = 0$, λ is a nuclear scale length, $v_c$ is the circular velocity, and $G$ is the gravitational constant. Our initial condition is an infinitesimal disk, but initial velocities are three dimensions, so that the disk rapidly acquires a finite thickness. The initial velocity dispersion is chosen to be 20% of the circular velocity (and locally isotropic), with is sufficient to suppress the usual Toomre instability [10] except in the innermost regions not of interest for these simulations.

The disk extends out to $r = 100$ units (so that edge effects are not important at the shear layer), while the boundary between the rotating and counter-rotating flows is a 2 unit wide annulus centered at $r = 25$ units. In the boundary region there is a linear transition between the rotating and counter-rotating flows. The accreting flow, exterior to the bounding annulus rotates in the CW sense, while the galaxy interior to the bounding annulus rotates in the CCW sense. The orbital period at $r = 25$ is 1.18 GADGET time units. A halo of mass equal to the disk has been included to suppress the bar-forming instability.

---
[5] *Mathematica* is a registered trademark of Wolfram Research, Inc.

## 4. Simulation showing development of instability in time

In figure 2 we show the progression of the simulation at selected times. (Times are indicated in each pane in GADGET units.) In the online version of this paper, the particles in the accreting flow are shown in red, which will show as grey in the print version of the paper. The particles in the accreting galaxy are shown in black.

By a time of 7.5 a well-developed spiral disturbance appears in the accreting galaxy that has a trailing wave pattern, while a spiral disturbance is beginning to develop near the shear layer in the accreting flow. These patterns initially have a very high azimuthal mode number that decreases as simulation time advances. The spiral opening angles in both the interior and exterior flows open up in time. At time 40, a ring composed mostly of accreting material has developed with spiral disturbances penetrating the ring to make material clumps. Accreting material is transported interior to the shear layer as the instability develops, but some of the accreting material is also scattered to large radii, presumably due to gravitational scattering by clumps of material in the ring and spiral structures.

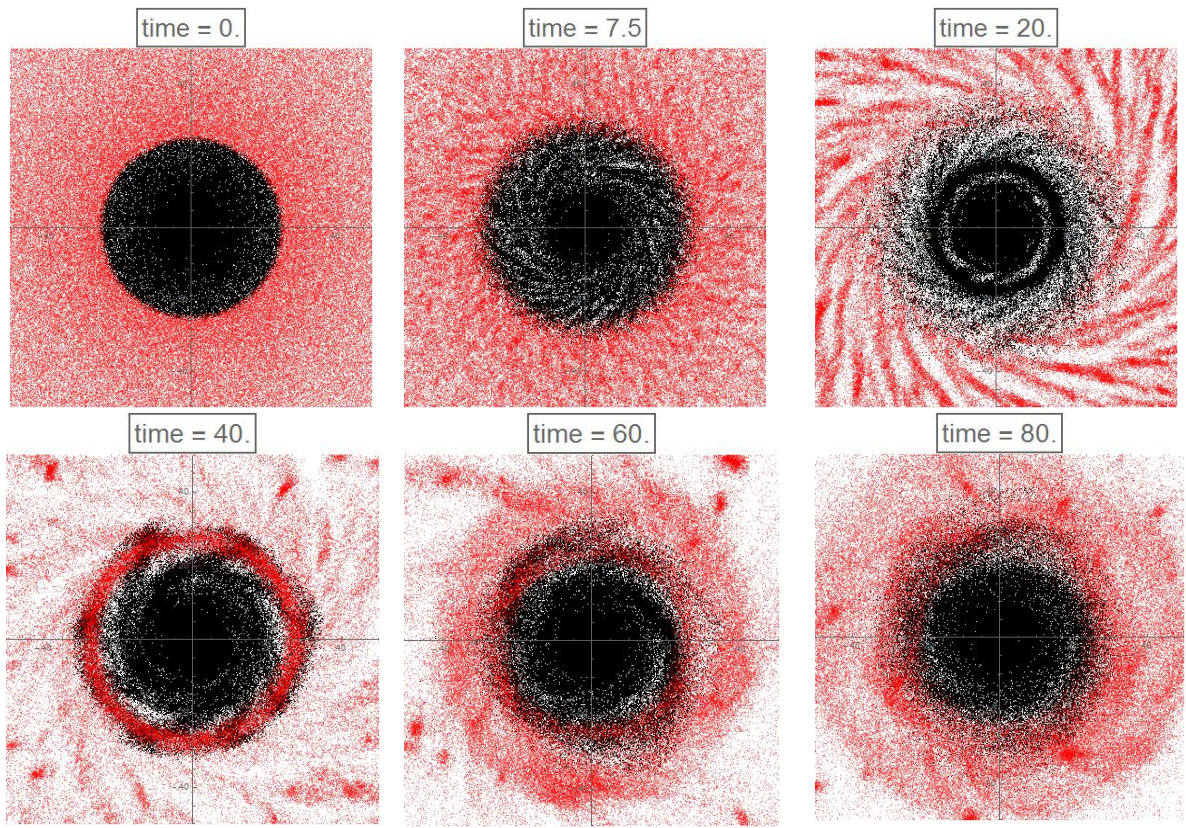

**Figure 2.** Selected images in the simulation showing the development of the instability in time. Times are in GADGET units. Accreting material is moving in the CW sense, while the accreting galaxy rotates in the CCW sense. A ring composed mostly of accreting material develops in the fourth pane, which shows clumping at a high azimuthal mode number. Accreting material is transported both to larger and smaller radii relative to the shear layer radius. The orbital period at $r = 25$ is 1.18 GADGET time units.

The two-stream instability for galactic accretion flows [4] was originally directed towards understanding flows with relatively small azimuthal mode number. Motivated by these initial simulation results, work is underway to extend that theory to higher azimuthal mode numbers so as to make detailed comparisons between the theory and simulations, as well as with observational results.

**5. Azimuthal mode analysis**
It is apparent by naked eye examination of the frames in figure 2 that the instability has a comparatively high mode number, which is changing in time. However, it is also quite likely that different modes are present at any one time. Quantifying the azimuthal mode spectrum will be important in making connections between these simulations and the extensions of the theory under development.

We chose a simulation frame at a time of 32, when the ring structure is well-developed (as shown in the left-hand side of figure 3). We collected stars within an annulus in the simulation and collected them in 128 bins based on their angular coordinates. We performed a Fast Fourier Transform of this binned angular data and then found the complex amplitude squared of each bin of the transform to obtain the angular power spectrum of the instability in the annulus. In the top rightmost bin of figure 3, we show the angular mode spectrum for the annulus between $r = 24$ to $r = 26$. Note the prominent peak in the spectrum for a mode number of 8 to 10. The pane below this was generated from the annulus from $r = 22$ to $r = 24$, i.e. interior to the prominent ring. The peak at mode numbers 8 to 10 is much reduced and the mode power can be seen to be significantly less. Further studies will determine the development of the azimuthal mode spectrum in time and radius.

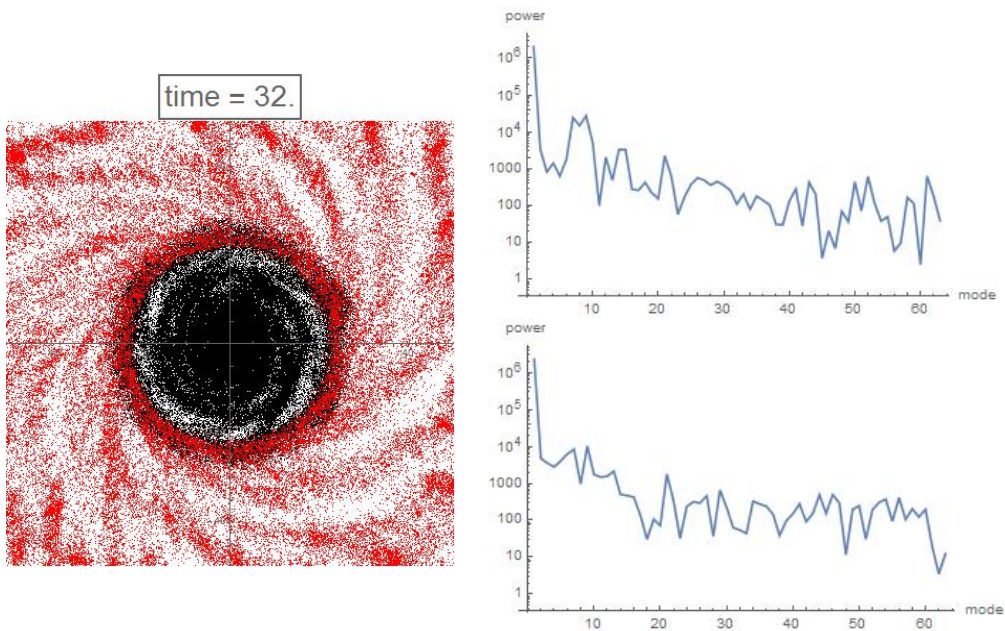

**Figure 3.** Azimuthal mode spectra for the simulation at time 32. The upper spectrum was collected in the annulus from radius 24 to 26, while the lower spectrum was collected from radius 22 to 24, interior to the prominent ring. Azimuthal mode numbers of 8 to 10 are important in the ring structure.

## 6. Conclusions
We list here the primary conclusions of this paper:

- Adjacent counter-rotating regions are unstable and interact by the two-stream instability.
- The instability has a high azimuthal mode number.
- A ring of material develops which contains particles with both senses of rotation.
- Clumps develop that are reminiscent of the star forming regions of NGC4862.
- Significant radial transport of accreting material occurs, both inward and outward.

**Acknowledgments**

We thank Professor Neil Comins for useful discussions. We also wish to acknowledge a generous allocation of computer resources on the Massachusetts Green High Performance Computing Center from Boston University's Research Computing Services and the Center for Computational Science.